%% file: curvapot.tex
\documentclass[dvips,11pt,a4paper]{article}

\usepackage[T1]{fontenc}         
\usepackage[latin1]{inputenc}
\usepackage{graphicx}
\usepackage{amsmath}
\usepackage{bm}
\usepackage{tikz}
\usepackage{color}
\usepackage{helvet}
\usepackage{courier}
\usepackage{makeidx}
\usepackage{footmisc}
\usepackage{type1cm}         
\usepackage{textcomp}
\usepackage{subfig}
\usepackage{enumitem,xcolor}    
\usepackage{yhmath}

\DeclareCaptionLabelSeparator{colon}{. }
\usepackage[labelfont={bf},labelsep={colon}]{caption}

\DeclareMathAlphabet\mathbfcal{OMS}{cmsy}{b}{n}

\begin{document}
\thispagestyle{empty}
\bibliographystyle{plain}



\include{article}


\end{document}

%% file: article.tex
\begin{center}
\textcolor{blue}{\Large \bf Prime Role of the Curvature of Potentials in Physics; Application to Inertia}
\end{center}

\author{Jean-Paul Caltagirone}

\begin{center}
Universit{\'e} de Bordeaux  \\
   Institut de M{\'e}canique et d'Ing{\'e}ni{\'e}rie \\
   D{\'e}partement TREFLE, UMR CNRS $n^o 5295$ \\
  16 Avenue Pey-Berland, 33607 Pessac Cedex    \\
\textcolor{blue}{\rm calta@ipb.fr}
\end{center}

\begin{abstract}
The curvature of the inertial or gravitational  potentials defined as a Hodge-Helmholtz  decomposition of acceleration into an irrotational and a solenoidal components, enable to federate certain domains of macroscopic physics. 
After two verifications in physics, one on the calculation of the curvatures for capillary effects and the second on the deflection of light by a gravitational effect, the concept of curvature of the potential is applied to the inertia. The physical analysis of each of the contributions of the decomposition is carried out on a classic example of fluid mechanics, the backward-facing step flow where inertia plays a preponderant role on the recirculation length. 
\end{abstract}

{\bf keyword}

Discrete Mechanics; Acceleration Conservation Principle; Hodge-Helmholtz Decomposition; Inertia; Navier-Stokes equations; General Relativity





\textcolor{blue}{\section{Introduction} }

The notion of curvature is important in all areas of physics, from the curvature of a surface in differential geometry to the curvature of space-time in general relativity. If the classical interpretation is trivial the physical phenomenon remains complex, the historical explanations of d'Alembert, Newton or Mach show that if the result is acquired the deep nature of this phenomenon remains mysterious.

Until today the inertia remains attached to the mass, without mass there can be no inertia. In relativity we define the mass of the photon with respect to its momentum in order to give it energy. The point of view developed in discrete mechanics is based on the contrary on the independence of the inertia with respect to the mass. The inertia of a material medium or massless photon exists, it is only related to its velocity by attributing to material velocity a meaning different from the celerity of the external environment, matter or vacuum. If, as in continuum mechanics, acceleration is replaced by the notion of force, mass becomes a simple multiplier for defining the force of inertia. 

Discrete Mechanics \cite{Cal19a} is a physical model based on Galileo concepts of relativity and equivalence between inertial and gravity effects. The fundamental law of dynamics is recovery by adopting this principle of equivalence and by suppressing the mass, the law of motion becomes an equality between accelerations, that of the material medium or of the particle with or without mass and those which are applied to it. The derivation of the equations of the discrete mechanics leads immediately to a Hodge-Helmholtz (HH)  decomposition of the acceleration in two orthogonal terms characterizing the effects of compressibility and rotation. The discrete geometric topology used corresponds to a local frame of reference. 

The idea developed here concerns the role of the curvature of some of the potentials in physics. A first verification is to find the result predicted by A. Einstein by the general relativity of the deflection of light by the sun due to space-time curvature of induced by it \cite{Wil18}. The second verification relates to capillary effects whose intensity is directly related to the main curvature of the interfaces between two fluids.

The notion of curvature of potentials is then extended to inertia by generalizing Bernoulli's law to flows of any dimension of space. The inertial acceleration, like the others, is due not to one but to two orthogonal contributions, one divergence-free and the other irrotational, the gradient and the dual curl of the inertial potential $\phi_i = \| \mathbf V \|^2/2$.

\textcolor{blue}{\section{Discrete formulation} }

\textcolor{blue}{\subsection{Bases of discrete mechanics} }

\begin{figure}[!ht]
\begin{center}
\includegraphics[width=5.4cm,height=4.cm]{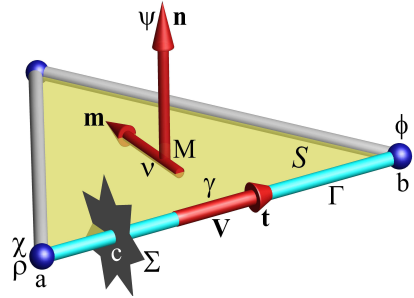}
\caption[Elementary geometrical structure of discrete media mechanics]{\it  Elementary geometrical structure of discrete media on the local frame of reference $( \mathbf m, \mathbf n, \mathbf t)$: three straight $\Gamma$ edges delimited by dots define a planar face $\mathcal S$. The unit normal vectors $\mathbf n$ to the face and the vector carried by $\Gamma$ are orthogonal, $\mathbf t \cdot \mathbf n = 0$. The edge $\Gamma $ can be intercepted by a discontinuity $\Sigma$ located in $c$, between ends $a$ and $b$ of $\Gamma$. $\phi$ and $\mathbf \Psi$ are the scalar and vector potentials respectively. }
\label{discreet}
\end{center}
\end{figure}

Discrete mechanics reformulates the fundamental law of dynamics from Galileo's only two intuitions, the Weak Equivalence Principle (WEP) \cite{Wil18} and the notion of relativity. The concepts of continuous medium, one-point derivation and analysis, etc. are abandoned in favor of a geometric description where, for example, velocity and acceleration are represented by a quantity $\mathbf V$ or $\bm \gamma$ constants on a $\Gamma$ oriented rectilinear segment delimited by its two ends $a$ and $b$ (figure \ref{discreet}). The local frame of reference $(\mathbf m, \mathbf n, \mathbf t)$ allows the different vectors, velocity, acceleration, curl and so on to be expressed. The derivation of the discrete mechanics is developed in another work \cite{Cal19a}.

The principle of equivalence applied to the fundamental law of classical mechanics $\mathbf F = m \: \bm \gamma$ with $\mathbf F = m \: \bm h$;  in the case of a force of gravitational origin, makes it possible to write $m \: \mathbf g = m \: \bm \gamma$ and to suppress the mass since it is the same on both sides of the equality, even if it is zero; contrary to the theory of relativity, which defines a mass at rest and another moving in the context of a change of reference, discrete mechanics defines only a local reference frame and all the interactions are of cause and effect.

There are two motivations for the suppression of mass in the equation of motion. The first relates to the gravitational attraction of light for a massless particle, the photon; the fundamental equation of dynamics will thus also be extended to media of zero mass, for example vacuum. The second argument relates to the dimensions with the physical magnitudes: all those which involve the mass are at the order one; it is then possible to redefine analogous quantities per unit mass. Finally, since gravitation is only a special case of acceleration, the quantity $\bm h$ will become a generic acceleration; the fundamental equation of discrete mechanics can therefore be written:
\begin{eqnarray}
\displaystyle{ \bm \gamma = \bm h  } 
\label{loiphys}
\end{eqnarray}
where $\bm h$ is the sum of the forces per unit mass, an acceleration.

This law (\ref{loiphys}) is expressed by `` the acceleration of a particle or a material medium with or without mass is equal to the sum of the accelerations which are applied to it ''. It will always be possible to return to quantities depending on the mass, energy for example; that we write $e = m \: c^2$ or $\phi = e / m = c^2$ does not change the physics.

The acceleration $\bm \gamma$ of a particle, a flux of particles or a material is considered in discrete mechanics as an absolute quantity attached to the single rectilinear segment $ \Gamma $; the acceleration $\bm h$ is the sum of the external accelerations defined similarly on the segment $\Gamma$. For acceleration, the principle of vector summation in the mathematical sense applies without reserve, whatever the velocity. Besides, the velocity $\mathbf V$ will only be a secondary quantity and the filtering of a uniform velocity superimposed on $\mathbf V$ will allow the principle of Galilean relativity to be completely satisfied.

The decomposition of a vector into an irrotational component and a solenoidal component will be postulated and applied to the acceleration $ \bm \gamma $:
\begin{eqnarray}
\displaystyle{ \bm \gamma = - \nabla \phi + \nabla \times \bm \psi  } 
\label{dechh}
\end{eqnarray}
where $\phi$ is the scalar potential and $\bm \psi$ the vector potential of the acceleration. It is always possible to define the potentials of any quantity, velocity for example, but only acceleration, an absolute quantity in a local reference frame, will have a physical reality. The potentials of (\ref{dechh}) are only defined to harmonic functions and depend on the boundary conditions. As the decomposition does not bring anything, the potentials $\phi$ and $\bm \psi$ must be expressed according to the same variable and it will be the local velocity $\mathbf V$ even if a certain amount of work remains; only the variables $(\bm \gamma, \phi, \bm \psi)$ will define the evolutions of the considered system.

\textcolor{blue}{\subsection{Discrete motion equation} }

The discrete motion equation is derived from the conservation equation of acceleration (\ref{dechh}) by expressing the deviations of potentials $\phi$ and $\bm \psi$ as a function of velocity $\mathbf V$. These ``deviators'' are obtained on the basis of the physical analysis of the storage-destocking processes of compression and shear energies; the first is written as the divergence of velocity and the second as a dual curl of velocity. The physical modeling of these terms is developed in a book devoted to discrete mechanics \cite{Cal19a}.

The vectorial equation of the movement and its upgrades is written:
\small
\hspace{-3.mm}
\begin{eqnarray}
\left\{
\begin{array}{llllll}
\displaystyle{\! \bm \gamma = - \nabla \! \left( \phi^o - dt c_l^2  \nabla \cdot \mathbf V \right) + \nabla \! \times  \! \left( \bm \psi^o - dt  c_t^2  \nabla \times \mathbf V \right) \!  + \! \bm h_s } \\  \\
\displaystyle{ \alpha_l \: \phi^o - dt \: c_l^2 \: \nabla \cdot \mathbf V \longmapsto \phi^o  } \\ \\
\displaystyle{  \alpha_t \: \bm \psi^o - dt \: c_t^2 \: \nabla \times \mathbf V \longmapsto \bm \psi^o } \\ \\
\displaystyle{\mathbf V^o + \bm \: \gamma \: dt \longmapsto \mathbf V^o  } 
\end{array}
\right.
\label{discrete}
\end{eqnarray}
\normalsize

Quantities $\phi^o$ and $\bm \psi^o$ are the equilibrium potentials, the same ones that allow the equation to be satisfied exactly at discrete instants $t$ and $t + dt$; $c_l$ and $c_t$ are the longitudinal and transverse celerities, intrinsic quantities of the medium that may vary according to physical parameters.
The terms $dt \: c_l^2 \: \nabla \cdot \mathbf V$ and $dt \: c_t^2 \: \nabla \times \mathbf V$ are respectively the deviations of the compression and shear effects. The right-hand side is thus composed of two oscillators where $\phi^o$ and $\bm \psi^o$ which represent energies per unit mass exchange these with their respective deviations. The two terms in gradient and in dual curl are orthogonal and cannot exchange energy directly; if an imbalance due to an external event occurs on one of these effects, then the acceleration is changed and the energy is then redistributed to the other term. The acceleration $\bm h_s$ represents gravity or any other source quantity and will also be written in the form of a HH decomposition.

The physical parameters $\alpha_l$ and $\alpha_t$ are the attenuation factors of the compression and shear waves. They also depend only on the medium considered, for example a Newtonian fluid retains the shear stresses only for very weak relaxation time constants, of an order of magnitude of $10^{- 12} \: s$ and the factor $\alpha_t$ can be taken as zero. The updating of potentials at time $t + dt$ is thus affected by these coefficients ranging between zero and unity. The velocity and possibly the displacement $\mathbf U$ are updated in turn. When the density is not constant, it is also updated using the conservation of mass in the form $\rho = \rho^o - dt \: \rho^o \: \nabla \cdot \mathbf V$. This quantity is only a function of the divergence of velocity.

It should be noted that $\phi^o$ and $\bm \psi^o$ are energies per unit mass, with each of the two terms reflecting the behavior of the medium or particle with respect to longitudinal and transverse waves. The term $\left (\phi^o - dt \: c_l^2 \: \nabla \cdot \mathbf V \right)$ is the longitudinal wave where $\phi^o$ accumulates the compression energy contained in the term deviator or restores it over time with celerity $c_l$. Similarly $\left (\bm \psi ^ o - dt \: c_t ^ 2 \: \nabla \times \mathbf V \right)$ is the oscillator corresponding to the velocity of the transverse waves $c_t$. The equations of physics are generally in a scalar, vectorial or tensorial form, but even in the case of vector or tensor equations, such as the Maxwell \cite{Max65} equations, they are decoupled which inhibits any instant energy transfer between the two types of waves; for example, the electric field $\mathbf E$ and the magnetic field $\mathbf B$ are coupled via second members. When a vector potential is defined, it is that of the magnetic field to impose $\nabla \cdot \mathbf B = 0$ and not the potential of the acceleration. The absence of nonlinear terms of inertia within Maxwell's equations is also a fundamental difference with discrete mechanics.

The time interval between two observations of the physical system in equilibrium is arbitrary, for example $ dt = 10^{20} \: s$ to obtain a stationary state at $dt = 10^{-20} \: s$ for the study in direct simulation of $\gamma$ rays. The solution does not depend on this parameter if it is adapted to the physics being mimed, in the general case that the solution is of order two in space and time. Since all the terms are implicit on the variable velocity, or linearized as the terms of inertia, the system (\ref{discrete}) is particularly robust.

The system (\ref{discrete}) is generic and allows us to mimic the phenomena of solid mechanics, fluid mechanics, electromagnetism or optics without any modification; the properties, $(c_l, c_t, \alpha_l, \alpha_t)$ are of course specific to the media considered and may vary over several tens of orders of magnitude. For example, it is the product $dt \: c_l^2$ which is significant of the phenomenon considered. Thus, for a gas flow, the time to be retained is $dt \approx 1 / c_l^2 \approx 10^{-5} \: s$, whereas for visible light it is necessary to keep a low lapse of time $dt \approx 1 / c_l^2 \approx 10^{-17} \: s$.
The resolution of this system can be achieved with the usual variables of each of the domains concerned or in terms of potentials.
It is possible to return at any time to the classical variables of fluid and solid mechanics or electromagnetism, but returning to the variables of the domains represented is not required. The solution of a problem in fact depends only on the variables $(\mathbf V, \phi, \bm \psi)$ and of course on the intrinsic properties of the environments studied; we note that the latter are often involved in the form of products of conventionally used properties.

\textcolor{blue}{\section{Curvatures of potentials} }

Relation (\ref{discrete}) looks like a classical Hodge-Helmholtz decomposition \cite{Ran19} of a vector $\bm h$ in the form $\bm h = - \nabla \phi + \nabla \times \bm \psi$ but when written in this way, the potentials $\phi$ and $\bm \psi$ can only be obtained at a gradient or at a divergence of arbitrary functions; they also depend closely on the boundary conditions of the problem \cite{Bha12}. The application of the divergence operator leads to $\nabla \cdot \bm h = - \nabla^2 \phi$ and the curl allows us to remove the gradient term $\nabla \times \bm h = \nabla \times \nabla \times \bm \psi$.
The introduction of the $\mathbf V$ variable related to acceleration $\mathbf V = \mathbf V^o + \bm \gamma \: dt$ makes it possible to transform the problem of the extraction of the potentials of the vector $\bm h$ into an unsteady problem that is written as $\bm \gamma = - \nabla \phi + \nabla \times \bm \psi - \bm h$; the acceleration $\bm \gamma$ and the velocity $\mathbf V $ are from zero to convergence at the end of the incremental process and we obtain the desired decomposition $\bm h = - \nabla \phi^o + \nabla \times \bm \psi^o$. The relation (\ref{discrete}) then becomes a real extractor of the irrotational and solenoidal components of a vector obtained implicitly.

The general form of the external accelerations $\bm h$ applied to a particle or a material medium therefore corresponds to a HH decomposition into a free divergence contribution and a irrotational one. In fact, because of the orthogonality of the main directions of the local coordinate system and the uniqueness of the physical potentials, the vector potential is none other than the association of the scalar potential with the normal of the primal planar surface $\mathbf n$, $\bm \psi = \phi \: \mathbf n$. The external acceleration will be of the form $\bm h = - \nabla \phi + \nabla \times (\phi \: \mathbf n)$.
\begin{table}[!ht]
\begin{center}
\begin{tabular}{|c|c|c|c|c|c|}   \hline
     interaction     &   $\phi$      \\ \hline  \hline 
 gravitation            &  $\displaystyle{ \phi_g = \frac{\mathcal G \: M }{ r } } \vspace{0.5mm} $   \\ \hline \vspace{0.5mm}
 capillary effects      &  $\displaystyle{  \phi_c = \frac{\sigma }{ r } }$    \\ \hline  \vspace{0.5mm}
 electromagnetism & $\displaystyle{  \phi_e =  \frac{ k_c \: q }{  r } }$    \\ \hline 
 inertia     & $\displaystyle{  \phi_i = \frac{  \| \mathbf V \|^2 }{ 2 } }$    \\ \hline 
\end{tabular}
\caption[Correspondence]{ \it Scalar potentials corresponding to different macroscopic interactions; $ \mathcal G $ is the universal constant of gravitation, $M$ the mass of the considered body, $r$ the distance, $\mathbf V$ the velocity of the medium, $\sigma$ the surface tension per mass unit , $k_c = 1 / (4 \: \pi \: \varepsilon)$ is the Coulomb constant where $\mu_0$ is the permittivity of the vacuum and $\varepsilon_0 = 1 / (\mu_0 \: c_o ^ 2)$ , $q$ the electric charge. }
\label{potar}
\end{center}
\end{table}

The table (\ref{potar}) gathers the interactions coming from well-known scalar potentials.
Some interactions not present in the table (\ref{potar}) and are taken into account directly by the equation of motion (\ref{discrete}). This is the case for the electromagnetic interaction where the law of discrete motion replaces the Maxwell \cite{Max65} equations by including the terms of inertia. This is also true for other fields of physics such as solid and fluid mechanics. This equation (\ref{discrete}) unifies these different fields of physics by introducing the potentials of acceleration. It should be noted that this law has only two fundamental units, a length and a time. The introduction of additional potentials, gravitation and molecular interaction etc. makes it possible to extend the desired unification.
The dual action principle that describes each fundamental interaction as the superimposition of a direct action represented by a gradient and another fixed by the dual curl is defined from the sole scalar potential $\phi$. These two components are the elements of the curvature of this potential.

Over the last centuries, profound differences in the interpretation of elementary phenomena, gravitation, inertia and mass, etc. have been revealed. Questions about the real nature of these concepts are still relevant, and possible unification is only possible if these questions find a global answer.

In mechanics, the notion of curvature is essential, being involved in many effects such as inertia, gravitation and capillarity etc. The notions of principal curvature, mean curvature, Gaussian curvature, etc. were first introduced in planar geometry intuitively and then in space in a rigorous mathematical framework. The curvature is defined by the Riemann curvature tensor; this corresponds to the acceleration with which two neighboring geodesics are separated on a curved space. In general relativity, gravitation is considered as a manifestation of the curvature of space-time.

 Two simple examples make it possible to show the generality of the proposed idea. The first  on small scale interactions related to capillarity  and the second on the deviation of light by the sun to find one of the results of general relativity. The main application of the presented concept is related to the physical nature of inertia.

\textcolor{blue}{\subsection{Capillary effects in fluid mechanics} }

{\it Capillarity is the curvature of the capillary potential $\phi_c = \sigma / r$. }
\vspace{2.mm}

Capillary effects due to interactions at the molecular scale can also be formulated in the same way by introducing the capillary potential $ \mathbf \phi_c = \sigma / r $ where $ \sigma $ is the surface tension per unit mass and $ r $ the radius of curvature, with each of these two physical parameters having a longitudinal contribution and a transversal one, like the two main curvatures of a surface in a three-dimensional space.
\small
\begin{eqnarray}
\hspace{-3.mm}
\left\{
\begin{array}{llllll}
\displaystyle{ \! \bm \gamma_c = - \nabla \! \left( \phi^o_c - \! \frac{dt  \sigma }{r}  \nabla \cdot \mathbf V \right) \! + \nabla \! \times \! \left( \bm \psi^o_c - \! \frac{dt  \sigma }{r}  \nabla \times \mathbf V \right) \! } \\  \\
\displaystyle{ \alpha_l \: \phi^o_c - \frac{dt \: \sigma }{r} \: \nabla \cdot \mathbf V \longmapsto \phi^o_c  } \\ \\
\displaystyle{  \alpha_t \: \bm \psi^o_c -\frac{dt \: \sigma }{r} \: \nabla \times \mathbf V \longmapsto \bm \psi^o_c } \\ \\
\displaystyle{\mathbf V^o + \bm \: \gamma \: dt \longmapsto \mathbf V^o  } 
\end{array}
\right.
\label{capill}
\end{eqnarray}
\normalsize

This system of equations can be used to calculate the only capillary effects on a surface where potential $\phi_c$ applies for example to the surface of a drop separating two liquids or on that of a bubble. However, the movement of an interface is in most cases related to the viscosity and density properties of the media. The equilibrium scalar potential $\phi^o$ can then be introduced into the equation of motion in the form $ - \nabla \phi_c + \nabla \times (\phi_c \: \mathbf n) $. In both cases, the capillary effects are implicitly taken into account by the divergence and curl of the velocity.

Since the unit vectors $ \mathbf t$ and $ \mathbf n$ of the discrete geometric structure are orthogonal, $\kappa_l$ and $\kappa_t$ represent the longitudinal and transverse curvatures; the sum $\kappa = \kappa_l + \kappa_t$ is the mean curvature. In the case of a sphere of radius $r$, the mean curvature is equal to $\kappa = 2 / r$ and in the case of a catenoid it is equal to zero.

In all the classical phenomena where the capillary effects are dominant, a capillary rise, a dynamic effect in two-phase flows, possible Marangoni-type effects, etc., the formulation also makes it possible to treat the movement of triple lines in a very natural way. The reference \cite{Cal19a} brings together numerous cases of two-phase problems where capillarity is dominant.

\textcolor{blue}{\subsection{Bending of light} }

{\it Gravitation is the mean curvature of the gravitational potential $\phi_g = \mathcal G \: M / r$.}
\vspace{2.mm}

The gravitational contribution to the acceleration of a particle or a material medium $\bm h_g$ is defined by $\phi_g = \mathcal G \: M / r$ where $\mathcal G$ is the universal constant of gravitation, $M$ the body mass and $r$ the distance from the point considered to the center of gravity of the body. In classical mechanics, only the contribution in $\nabla \phi_g$ is used, which induces  an error in the deviation of the trajectory of the light when it crosses the neighborhood of the sun.

The solution of the equations of classical mechanics can be directly sought in polar coordinates and, with a few small approximations, obtained analytically and is usually reproduced in university textbooks. A approximate value of deviation was obtained by Soldner in 1804 \cite{Sol04}. 
CM Will \cite{Wil88} asserts that H. Cavendish found the correct result well before J. Soldner, around 1784, considering that the photon trajectory ranged from $- \infty$ to $+ \infty$. Other hypotheses are advanced by considering that the light behaves in the same way as when it crosses a prism, by considering two refractions, one at the entrance of the prism and the other at the exit. Still others attribute the doubling of the value obtained by integrating the equations of classical mechanics taking into account the inertia of the photon but also that of the sun \cite{Dei16}. 

Consider first the so-called case of "Newtonian mechanics" where the equation of motion is written in the form:
\begin{eqnarray}
\displaystyle{ \frac{d \mathbf V}{d t} = - \frac{\mathcal{G} \: M_{\odot} }{r^2} \: \mathbf e_r  } 
\label{gravitee}
\end{eqnarray}
where ${\mathcal G} = 6.6738480 \: 10^{- 11} \: m ^ 3 \: kg ^{- 1} \: s ^{- 2} $ is the universal constant of gravitation, $ M_{ \odot} = 1.99 \: 10^{30} \: kg $ the mass of the sun and $ R = 6.95 \: 10^8 \: m $ its mean radius. The value of the deviation is equal to:
\begin{eqnarray}
\displaystyle{ \Phi = \frac{ 2 \: \mathcal{G} \: M_{\odot} }{V_o^2 \: R}   } 
\label{phi}
\end{eqnarray}

We note that the celerity of the Light, $c_0 = 2.99792458 \: 10^8 \: m s^{-1}$ is not present in this equation, which is natural in this simplified context.
The celerity $c_0$ is of course absent from this result, but assuming that the initial velocity $ V_o $ is equal to the celerity of the light, we find $\Phi = 4.250 \: 10^{- 6} \: rd $ is $ \Phi = 0.875 ''$ of arc.
The value measured by Eddington in 1919 \cite{Dys20} was $\Phi = 1.75 ''$ arc or exactly double.
The success of the theory of relativity was due mainly to the verification of Einstein's prediction established a few years before. Many other phenomena have been explained and verified since, for example the existence of black holes in the case of massive stars of small diameter; for the sun, the Schwarzschild radius $ R_s $ needed to capture light would be $ R_s = 2 \: \mathcal{G} \: M _{\odot} / c_0^2$ that is $ R_s = 2.95 \: km $, a value much lower than its radius.

This major result of general relativity is the opportunity to test discrete mechanics \cite{Cal19c} and its equation of motion (\ref{discrete}) outside its usual field of application.  The dual action hypothesis postulates that any scalar potential of physics $\phi$ is written as the sum of a gradient of the latter and a dual curl of a vector potential equal to $\bm \psi = \phi \: \mathbf n$.

The explanation proposed here on the problem posed is of a different nature: the deflection of light by the sun is due to a competition between the inertial and gravitational effects. A particle with a mass or massless has an inertia characterized by the inertial potential $\phi_i = \| \mathbf V \|^2/2$; if it is in a gravitational field it is subjected to an acceleration fixed by the potential $\phi_g$. The light is deflected in a single direction contained in a plane passing through the center of the sun but with an intensity that is related to the mean curvature of the gravitational potential.

Consider two vectors $\mathbf t$ and $\mathbf m$ corresponding to the principal directions on an equipotential surface $\phi_g$ such that $\phi_g = Cte$. These two vectors are orthogonal by definition $\mathbf t \cdot \mathbf m = 0$ and we show in differential geometry that the mean curvature $\kappa$ does not depend on the choice of $\mathbf t$ and $\mathbf m$ on the $\phi_g$ surface.

At first order, if the velocity of the particle is always equal to the celerity in vacuum $c_0$, the deviation is written as the ratio of the gravitational and inertial mean curvatures:
\begin{eqnarray}
\displaystyle{ \Phi \approx \frac{\kappa_g}{\kappa_i} = \frac{ \| - \nabla \phi_g + \nabla \times \bm \psi_g \|  }{ \| \nabla \phi_i \|} =  \frac{ 2 \: \phi_g}{ \phi_i}  = \frac{ 4 \: \mathcal{G} \: M_{\odot} }{c_o^2 \: R} } 
\label{deviation}
\end{eqnarray}

This is the result predicted by A. Einstein, measured by Eddington in 1919 and confirmed by more recent experiments. The Eulerian vision of the movement of a particle in the vicinity of a massive body leads to explicitly taking into account the inertia independently of the mass of the latter. The result is consistent with that of a Lagrangian vision given elsewhere \cite{Cal19a}.

The concept of curvature of space-time introduced in general relativity makes it possible to explain the deflection of the light by the sun, replaced here by that of the mean curvatures of the gravitational  and inertial potentials. If the result is the same, the formulation of the continuum equations of general relativity is much more complex and requires the use of geodesic calculations and curvature tensor.
The problem dealt with in this section corresponds to an unsteady mechanical equilibrium due to the curved  trajectory of the photon over a short period of time when it passes near the sun. The following case is stationary and the mechanical equilibrium is ensured in the long time by the scalar and vector potentials.

\textcolor{blue}{\section{On the nature of inertia} }

{\it Inertia is the curvature of inertial potential $\phi_i = \| \mathbf V \|^2 / 2$. }
\vspace{2.mm}

In continuum mechanics, the acceleration $\bm \gamma$, the material derivative of velocity, is written $ \partial \mathbf V / \partial t + \mathbf V \cdot \nabla \mathbf V $ or, using the Lamb vector, $\partial \mathbf V / \partial t - \mathbf V \times \nabla \times \mathbf V + \nabla \left (\| \mathbf V \| ^ 2/2 \right) $.  These terms of inertia were modeled within the framework of the continuum mechanics cannot be transformed into a HH decomposition, the Lamb term is not a curl one.

The physical idea is based on the following postulate: if a direct effect represented by the gradient of a scalar potential exists, here $\phi_i = \| \mathbf V \|^2/2$ the inertial potential,  then its dual effect represented by the dual curl of the vector $\bm \psi_i = \| \mathbf V \|^2/2 \: \mathbf n$ also exists.
 The acceleration is thus rewritten in the form:
\begin{eqnarray}
\displaystyle{ \bm \gamma = \frac{d \mathbf V}{d t} =  \frac{\partial \mathbf V}{\partial  t} + \nabla  \left( \frac{\| \mathbf V \|^2 }{2}  \right) - \nabla \times \left( \frac{\| \mathbf V \|^2 }{2} \: \mathbf n \right)  =  \frac{\partial \mathbf V}{\partial  t} + \bm \kappa_i  } 
\label{inertie}
\end{eqnarray}

First nonlinear term $(\nabla \phi_i)$ are already present in some laws, like the second law of Bernoulli, and the expression (\ref{inertie}) only generalizes the inertial effects in a three-dimensional space in a coherent way for the formalization of the equation of the movement itself. It physically represents the advection of the velocity by itself, independently of the mass of the considered particle, but this can be realized implicitly by integrating the terms of $\bm \gamma$ of the expression (\ref{inertie}) in the vector equation of the system (\ref{discrete}).

The classical version of the material derivative of velocity is thus conceptually revised. In fact it is a notion of a continuous medium which inhibits any possibility of finding an analytical representation of inertia in the form of a HH decomposition. The geometric character of the proposed discrete approach makes it possible to escape this restriction.

The origin of inertial forces has long been considered as being linked to gravitation, an idea suggested by the principle of equivalence between ``inert mass'' and ``gravitational mass'' introduced by Galileo. We find this interpretation in the last century, for example in A. Kastler \cite{Kas31} who wrote: ``the forces of inertia are forces of dynamic origin, produced by the acceleration of a body compared to the mass of the universe'' taking up again the principle of E. Mach. But the equality between accelerations does not mean that the nature of the phenomena is the same, they are simply equal. The Mach principle cannot be validated or reversed, it is simply inadequate; inertia exists with or without gravity, with or without mass. This is the curvature of the inertial potential $\phi_i$.

\textcolor{blue}{\subsection{Analysis of backward-facing step flow at $Re = 200$ } }

The  backward-facing step flow at $Re = 200$ was chosen because of its specific properties. First, the ratio quantized by the Reynolds number between inertial and viscous effects is sufficiently important to finely analyze the terms of inertia in the equation of motion. This flow is stationary and incompressible which facilitates the interpretation of phenomena. Moreover, the inertial effects are confined in the region of the nose of the step where the channel suddenly widens, the upstream and downstream flows correspond to non-inertial Poiseuille type velocity profiles.

Since the study by Armaly et al. \cite{Arm83} the publications devoted to this configuration are very numerous; the main influences of the backward-facing step characteristics are the Reynolds number, the ratio $ H / h $ between the height of the output and input  channels, the length of the output channel, the ratio  $l/h$ between the width in the direction orthogonal to the planar surface of the step and the height of it, see for example \cite{Bis04}.

The aim here is not to improve existing results on one aspect or another; it is to analyze as finely as possible the solenoidal and irrotational contributions of inertia on a fixed configuration. The ratio of the $H / h$ domain is taken as $2$ and the length of the output channel is large enough to avoid any disturbance of the condition at the downstream boundary. The main recirculation behind the step obtained by the simulation from the system of equations (\ref{discrete}) is entirely consistent with the results of \cite{Bis04}, it is $L / h = 4.95$. The simulation also shows a secondary vortex of Moffatt \cite{Mof64} in the lower corner of the step. The flow obtained is perfectly stationary and the order of magnitude divergence of the machine accuracy.

By taking again the definitions of the terms of inertia of the discrete mechanics by noting $\phi_i = \| \mathbf V \| ^ 2/2$ and $\bm \psi_i = \phi_i \: \mathbf n$ it is possible to focus on upstream and downstream flows where they are characterized by horizontal velocity profiles of the form $\mathbf V = (1 - y^2 / d^2) \: \mathbf e_x$ centered on $y = 1.5$ for the input and $y = 0$ for the output with respectively $d = h / 2$  and $d =  1$. Upstream and downstream the inertia must be zero because $\mathbf V \cdot \nabla \mathbf V = 0$. In discrete mechanics the two contributions $\nabla \phi_i$ and $\nabla \times (\phi_i \: \mathbf n) $ are not zero separately but the difference between the two is nil, $ \nabla \phi_i - \nabla \times (\phi_i \: \mathbf n) = 0$, because they are equal.

\begin{figure}[!ht]
\begin{center}
\includegraphics[width=5.5cm]{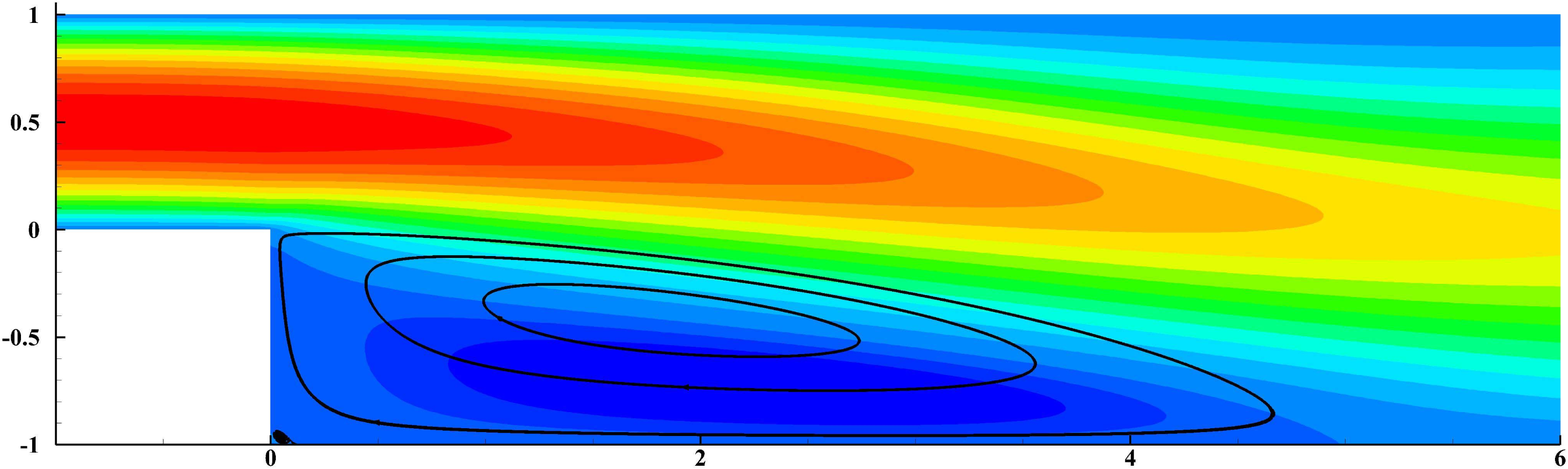}
\includegraphics[width=5.5cm]{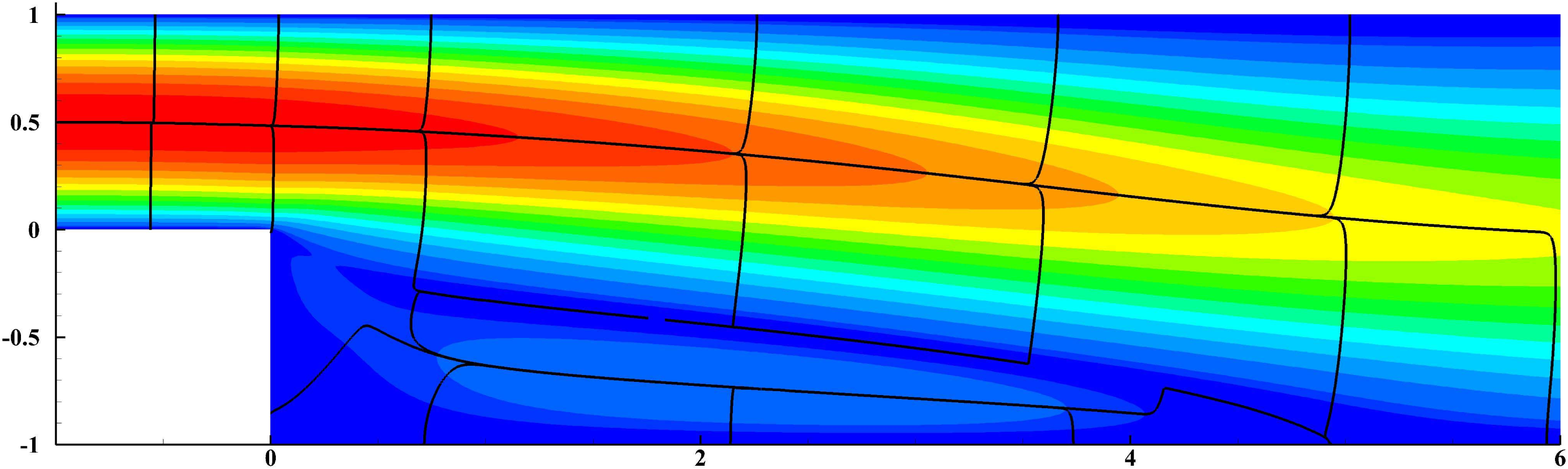}
\includegraphics[width=5.5cm]{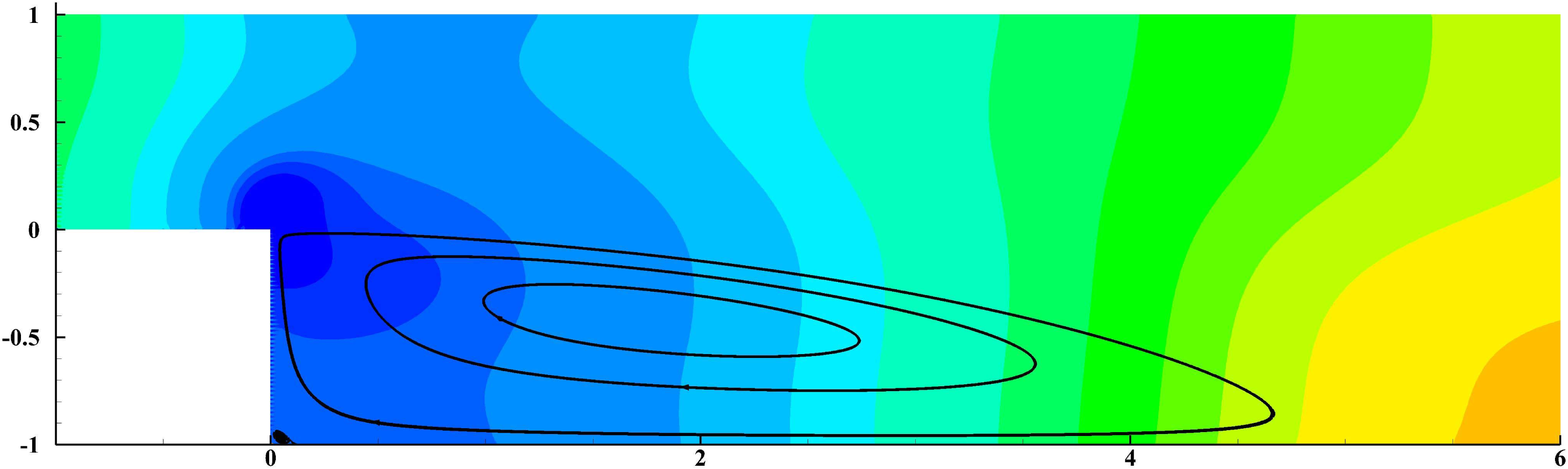}
\includegraphics[width=5.5cm]{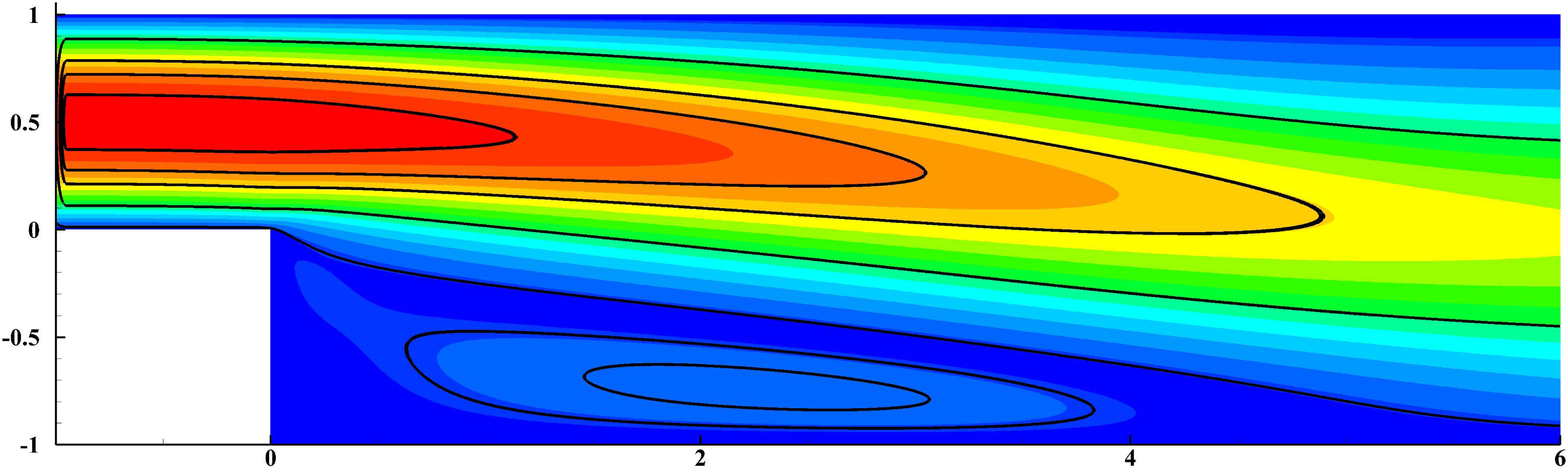}
\includegraphics[width=5.5cm]{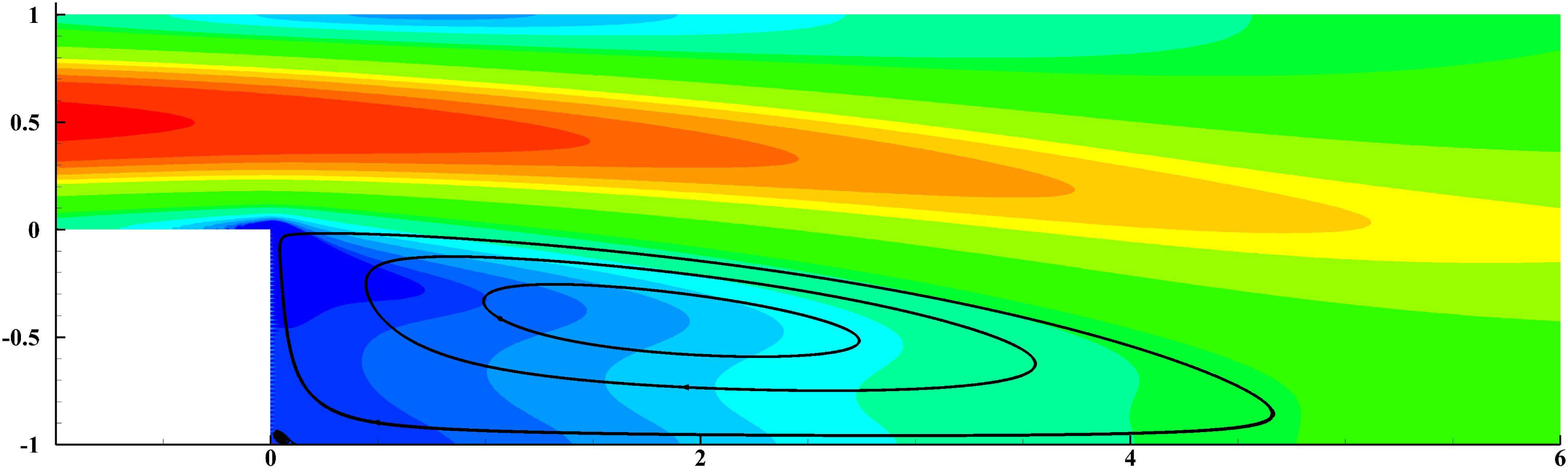}
\includegraphics[width=5.5cm]{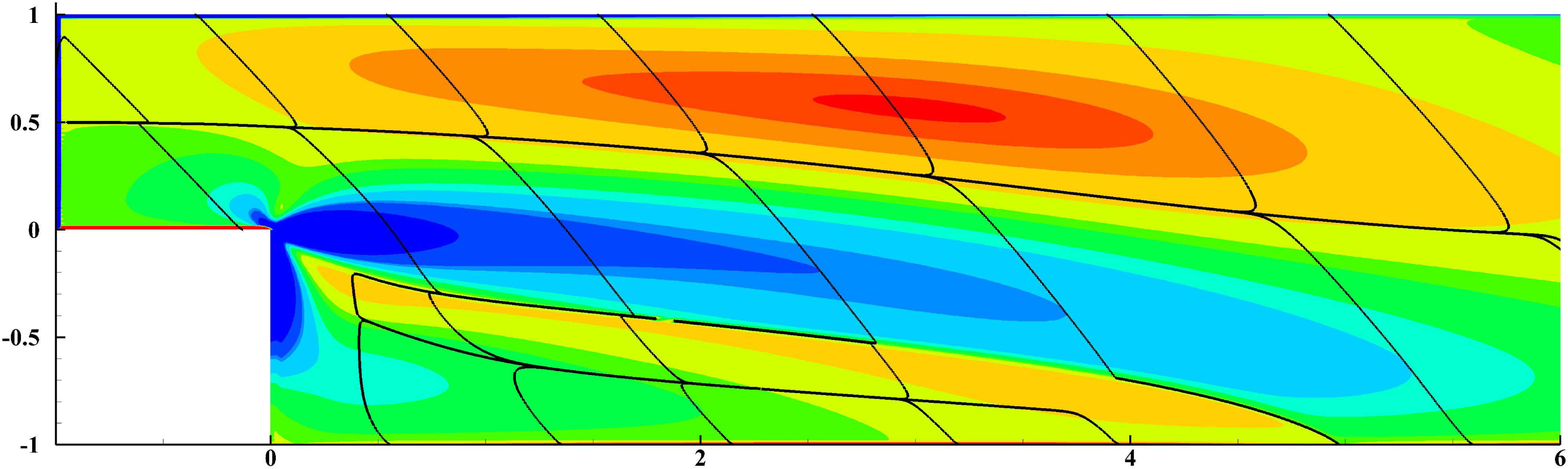}
\caption{\it Laminar backward-facing step flow at  $Re =  200$. On the left side and from top to bottom the component along $x$ of the velocity, the potential $\phi^o$ and the potential $\phi_B$ superimposed a few stream lines  in the recirculation. On the right is represented the scalar field $\| \mathbf V \|^2 / 2$ and the stream lines of the inertial terms $\mathbf R = 1/2 \: \nabla (\| \mathbf V \|^2)$ and $\mathbf G = 1/2 \: \nabla \times (\| \mathbf V \|^2 \: \mathbf n)$; the figure at the bottom right shows  the curvature $\kappa_i$ of inertia. }
\label{marche-200}
\end{center}
\end{figure}

Indeed the stationary solution corresponding to the flow of Poiseuille is equal to $\mathbf V = (1 - y^2) \: \mathbf e_x$ that is $\nabla \phi_i = [0, 9/2 \: (- y + y^3), 0]$ and $\nabla \times \bm \psi_i = [9/2 \: (- y + y^3), 0, 0]$. The components of $\nabla \phi_i$ and $\nabla \times \bm \psi_i$ do not relate to the same edge and the equality of the terms of inertia cannot be satisfied on each one of them. But the curvature $\kappa_i = - \nabla \phi_i + \nabla \times (\phi_i \: \mathbf n)$ must be evaluated at the barycenter of the cells or on the points of the primal geometric topology. For these two locations, the mean curvature of the inertia is indeed zero $\kappa_i = 0$. The figure (\ref{marche-200}) shows that the curvature becomes zero when one deviates from the widening zone of the channel upstream or downstream.

In the widening zone the fluid undergoes a downward deflection due to a marked depression which causes the flow to develop a main stationary recirculation and another of very low amplitude in the lower corner of the step. In this zone where the curvature $\kappa_i$ is important the terms of inertia are preponderant; it is negative below the median and positive dividing line above. Since the sum  is neither a gradient nor a curl current lines corresponding to this vector are not orthogonal nor aligned with the median trajectory. Similarly, the dividing line between the main flow and the recirculation defines two zones of positive and negative curvature.

In continuum mechanics the term of inertia is written in the form $\mathbf V \cdot \nabla \mathbf V $ it is not possible {\it a priori} to separate the solenoidal and irrotational contributions. The use of the Lamb vector poses other difficulties noted in the textbook \cite{Cal19a}. However, the solutions obtained on the velocity field $\mathbf V$ are strictly identical. The recirculation shown in the figure (\ref {marche-200}) is, for example, strictly in accordance with the results of the literature.

The differences between continuum mechanics and discrete mechanics are only revealed if gradient or curl operators are applied to the terms of inertia. The result is not the same and the difference is associated with the second invariant $I_2$ of the tensor $\nabla \mathbf V$. The use of second order tensors in continuous media mechanics is the source of this disagreement; $I_2 = 0$ is a compatibility condition to satisfy \cite{Cal19b} if the flow is incompressible.

From the physical point of view the inertia presented here as the curvature of the potential $\phi_i$ does not disagree with the classical definition "the inertia is a property of resisting by which every body endeavors to preserve its present state, whether it be of rest or of moving uniformly forward in a straight line"; in this case the curvature of the potential is zero. When the fluid must change direction, for example by the presence of an obstacle, it is subjected to an acceleration $\bm \gamma_i = - \nabla \phi_i + \nabla \times \bm \psi_i$. In one dimension of space only the contribution in gradient persists and one finds the potential of Bernoulli $\phi_B = \phi + \| \mathbf V \|^2/2$. It may be noted that between a point upstream of the flow around a circle and the break point the trajectory is straight and yet the inertia is not zero. The equilibrium of a material medium is governed by a set of accelerations, that of the medium and those which are applied to it from the outside, only the equation of the movement makes it possible to achieve this equilibrium.

\textcolor{blue}{\section{Conclusions} }

Discrete mechanics revisits the equations of mechanics from a different angle by abandoning a number of concepts based on a continuous approach. The Hodge-Helmholtz decomposition of all accelerations, inertial, gravitational, capillary, as well as longitudinal and transverse propagation, presented here as a postulate, is compatible with the results obtained by previous theories of mechanics. This does not mean less a radical change in point of view.
The concept of potential curvature makes it possible to understand and overcome a number of difficulties that arise in the search for the unification of the laws of physics, in fluid and solid mechanics, electromagnetism and optical physics, etc. 

The notion of inertia defined here as the curvature of the potential $\phi_i$ is general and applies for the material media as well as to the particles with or without mass. In discrete mechanics the equation of motion does not show mass and the energy per unit mass $e / m$ is simply the equilibrium scalar potential $\phi^o$; from then on, the photon possesses an inertia and its setting in motion up to the celerity of light $c_0$ depends on the acceleration that is applied to it. Reasoning in terms of force rather than acceleration would immediately affect it to a velocity equal to $c_0$, which is the case in physics today.

The test case of the backward-facing step flow at $Re =  200$ allows a fine analysis of each of the terms of the equation of motion and in particular terms of inertia. The results obtained in discrete mechanics do not differ from those of the mechanics of continuous media where the term of inertia is written $\mathbf V \cdot \nabla \mathbf V$; the use of second-order tensors reduced by suitable operators to return to an equation of vector motion requires the use of compatibility conditions that are difficult to implement. Here the terms of inertia are written from the gradient and dual curl operators consistent with the rest of terms of the  motion equation. From the physical point of view, the HH decomposition of inertia provides a structural framework by defining the inertia as the curvature of inertial potential.



\bibliography{database}



